\documentclass[11pt,a4paper]{article}
\pdfoutput=1
\usepackage{authblk}
\usepackage[english]{babel}  
\usepackage{amssymb,amsmath}    
\usepackage{fancyhdr}       
\usepackage[section]{placeins} 
\usepackage{psfrag} 
\usepackage{verbatim} 
\usepackage{epsfig}
\usepackage{enumerate}   
\usepackage{graphicx} 
\usepackage{caption}
\usepackage[pdftex]{hyperref}       
\usepackage{graphics}
\usepackage{anysize}
\marginsize{2.5cm}{2.5cm}{1cm}{1cm} 
\usepackage{multirow}

\newcommand{\bmb}{\begin{bmatrix}} 
\newcommand{\bme}{\end{bmatrix}}   
\newcommand{\ppmb}{\begin{pmatrix}} 
\newcommand{\ppme}{\end{pmatrix}}   
\newcommand{\equb}{\begin{equation}} 
\newcommand{\eque}{\end{equation}} 
\newcommand{\equab}{\begin{eqnarray}} 
\newcommand{\equae}{\end{eqnarray}} 

\begin{document}

\title{The Frobenius group $T_{13}$ and the canonical see-saw mechanism applied to neutrino mixing}
\author{Christine Hartmann}
\affil{Niels Bohr International Academy and Discovery Center, Niels Bohr Institute, University of Copenhagen, DK-2100 Copenhagen, Denmark
\\ Email: chartmann@nbi.ku.dk}
\date{\today}


\maketitle

$\bold{Abstract}$

The compatibility of the Frobenius group $T_{13}$ with the canonical see-saw mechanism of neutrino mixing is examined. The standard model is extended minimally by introducing a family symmetry and three right-handed neutrinos. To fit experiments and place constraints on the possibilities, tribimaximal mixing is used as a guideline. The application of both a family symmetry group and the canonical see-saw mechanism naturally generates small neutrino masses. The various possibilities from combining these two models are listed. Enough constraints are produced to narrow down the parameters of the neutrino mass matrix to two. This is therefore a predictive model, where the physical neutrino masses and the allowed regions for neutrinoless double beta decay are suggested.

\section{Introduction}

It is now widely accepted that neutrinos oscillate and therefore must be massive. What has not yet been understood is why neutrinos have such small masses compared to charged leptons and quarks. From cosmology (the large scale structure and the cosmic microwave background), the sum of the mass eigenstates has been given an upper limit \cite{Fogli}:

\equb \Sigma m_i \leq 0.7 eV. \eque
From the endpoints of tritium single beta decay, the following upper limit on the neutrino mass associated with the beta decay has been obtained \cite{Krauss, Lobashev}:

\equb m_{\beta} = (|U_{ei}|^2m_i^2)^{1/2} \leq 2 eV. \eque
As a result, it has been suggested that neutrinos acquire masses in a different way than their cousins.

The experimentally measured squared mass differences are shown in Table \ref{table:expvalues}, with best fit values in addition to $2\sigma$ and $3\sigma$ intervals. The values are from global data with solar, atmospheric, reactor and accelerator experiments \cite{Tortola}. Since the sign of the difference between $m_2^2$ and $m_3^2$ is unknown, there exists two possible mass hierarchies: the normal hierarchy, where $| m_3 | \gg | m_2 | \simeq | m_1 |$, and the inverted hierarchy where  $| m_3 | \ll | m_2 | \simeq | m_1 |$.

\begin{table}[ht]
\centering      
\begin{tabular}{c | c  c  c}  
\hline                    
Parameter & Best fit &2 $\sigma$ & 3$\sigma$ \\   
\hline
\\
$\Delta m_{12}^2 [10^{-5} eV^2]$& $7.59_{-0.18}^{+0.20}$ &  7.24-7.99 & 7.09-8.19  \\ 
\\
\multirow{2}{*}{$|\Delta m_{31}^2|[10^{-3}eV^2]$} & $2.50_{-0.16}^{+0.09}$ &  2.25-2.68 &  2.14-2.76 \\
& -$(2.40_{-0.09}^{+0.08})$ & -(2.23-2.58) & -(2.13-2.67) \\ 
\\
\hline
\end{tabular} 
\caption{Recent experimental values of mass squared differences for the normal (inverted) hierarchy \cite{Tortola}.} 
\label{table:expvalues}  
\end{table} 
If neutrinos are Majorana, neutrinoless double beta decay could be detectable. The limit from neutrinoless double beta decay on the modulus of the (ee) element of the neutrino mass matrix is given by \cite{Avignone,Fogli}

\equb |m_{ee}| \leq 0.38 eV. \eque
Neutrino masses can be understood in several ways. One way is to introduce right-handed neutrinos creating Dirac masses from Yukawa type coupling with the Higgs doublet. Another way is to introduce lepton number violation, where neutrinos are assumed to be Majorana. A third possibility is a mixture of both. This is the option associated with the see-saw mechanism, which has proven to give naturally small masses to the neutrinos. These arise from a relation between the Majorana and Dirac masses, which will be reviewed in section 2.2 \cite{Models}.

This paper will follow up on a previous paper \cite{Frobenius}, where the various possibilities using the Frobenius group $T_{13}$ with neutrino mixing were studied. A minimal approach was used supplementing SU(2)$\otimes$U(1) with this family symmetry group. Tribimaximal mixing was then applied to guide the process. Since tribimaximal mixing still gives a very good fit with experiments at lowest order, this paper will also use this mixing matrix as a guideline. The group $T_{13}$ will then be applied to neutrino mixing, implementing the canonical see-saw mechanism in which right-handed neutrinos are introduced. The canonical see-saw mechanism was previously applied in the case of the non-Abelian symmetry group $A_4$ in a very similar way \cite{Hybrid, A4seesaw, Kovtun}. 

In this paper, exact tribimaximal mixing is obtained by using the Frobenius group together with the canonical see-saw mechanism. This leads to one specific allowed neutrino mass matrix that depends on only two parameters. From this and the known experimental values, predictions for the neutrino mass eigenvalues are proposed in each of the mass hierarchies. Values are suggested for the modulus of the (ee) element of the neutrino mass matrix which is proportional to the neutrinoless double beta decay rate. The validity of this model can thereby be tested in the near future. 

\section{Review}

\subsection{Neutrino mixing}

Neutrino oscillations require a mixing matrix that transforms the mass eigenstates into the flavor eigenstates as follows:

\equb \left( \begin{array}{ccc}
\nu_e \\ \nu_{\mu} \\
\nu_{\tau} \end{array} \right)= V \left( \begin{array}{ccc}
\nu_1 \\ \nu_2 \\
\nu_{3} \end{array} \right). \label{mix}\eque
This mixing matrix is given by:

\equb \nonumber
V =  \left( \begin{array}{ccc}
e^{i\kappa_1} & 0 & 0 \\
0 & e^{i\kappa_2} & 0 \\
0 & 0 & e^{i\kappa_3}  \end{array} \right) U \left( \begin{array}{ccc}
1 & 0 & 0 \\
0 & e^{i\Phi_1} & 0 \\
0 & 0 & e^{i\Phi_2},  \end{array} \right) \eque
where
\equb U =  \left( \begin{array}{ccc}
c_{12}c_{13} & s_{12}c_{13}  & s_{13}e^{-i\delta} \\
-s_{12}c_{23} - c_{12}s_{23}s_{13}e^{i \delta} & c_{12}c_{23} - s_{12}s_{23}s_{13}e^{i\delta} & s_{23}c_{13}\\
s_{12}s_{23} - c_{12}c_{23}s_{13}e^{i\delta} & -c_{12}s_{23} - s_{12}c_{23}s_{13}e^{i\delta} & c_{23}c_{13} \end{array} \right). \eque
The three $\kappa$ phases can be absorbed by rephasing the neutrino field \cite{Phenomenology}. For Majorana neutrinos, besides the Dirac phase $\delta$, there are two extra Majorana phases, $\Phi_1$ and $\Phi_2$. These are relative phases among the Majorana masses that are not observable in neutrino oscillations, as is well known.

Experimental measurements are well described by the tribimaximal mixing matrix \cite{Stancu, Scott}:

\equb U_{TB} = \left( \begin{array}{ccc}
\frac{2}{\sqrt{6}} & \frac{1}{\sqrt{3}} & 0 \\
-\frac{1}{\sqrt{6}} & \frac{1}{\sqrt{3}} & \frac{1}{\sqrt{2}} \\
-\frac{1}{\sqrt{6}} & \frac{1}{\sqrt{3}} & -\frac{1}{\sqrt{2}} \end{array} \right). \eque
However, this mixing matrix cannot be exact \cite{T2K} and can therefore only be a first approximation. It will be used as a guide for model building in this paper.

\subsection{The canonical see-saw mechanism}

The canonical see-saw mechanism, also known as the type I see-saw model, is currently the most investigated mechanism for understanding neutrino masses \cite{Minkowski, Glashow, Gell-Mann, Senjanovic}. When introducing right-handed neutrinos, the effective dimension 5 operator, resulting in neutrino masses, can be obtained from tree-level interactions:

\equb  \mathcal{O}_5 = \frac{1}{M}(\phi \tau_2 \psi_L)^T C^{-1} (\phi \tau_2 \psi_L), \label{dim5} \eque
with $\phi$ the Higgs doublet, $\psi_L$ the left-handed lepton doublet, $C$ the charge conjugation operator, and $M$ the cutoff scale of the effective field theory. The Lagrangian from which this operator is extracted consists of a coupling between the right-handed neutrinos, the left-handed neutrinos, and the standard sodel Higgs doublet, as well as a Majorana mass term for the right-handed neutrinos:

\equb \mathcal{L}^{\nu} =  \bar{\psi}_L\lambda_{\phi}\phi N_R + \frac{1}{2}N_R^TC^{-1}M_RN_R + h.c. \label{Lag} \eque
When the Higgs doublet $\phi$ acquires a vacuum expectation value $v$, under spontaneous symmetry breaking, this leads to a Dirac neutrino mass, $M_D = \lambda_{\phi} v$. The mass matrix $M_D$ could therefore be expected to be at the same order of magnitude as $M_q$ and $M_l$, the masses of quarks and charged leptons. 

A Yukawa coupling term is allowed between right-handed neutrinos and a singlet scalar field, $\chi$. The mass matrix of the right-handed neutrinos, $M_R$, therefore consists of a bare mass term $M$ and a mass term from the Yukawa coupling, $\lambda_{\chi}\langle \chi \rangle$ when the scalar singlet acquires a vacuum expectation value. The Majorana term is more explicitly given by

\equb \frac{1}{2}N_R^TC^{-1}M_RN_R = \frac{1}{2}N_R^TC^{-1}MN_R  + \frac{1}{2}\lambda_{\chi}N_R^TC^{-1}N_R\chi. \eque
These Majorana masses are not related to the electroweak symmetry breaking and there is no symmetry protecting them, so they are naturally of the order of the cut-off scale \cite{Models}. The dimension 5 operator is then produced, for example, when integrating out the very heavy right-handed neutrinos, $N_{Ra}$, which are singlets under SU(2)$\otimes$U(1). The $M$ in the denominator of equation (\ref{dim5}) is now the mass matrix of the very heavy right-handed neutrinos $N_{Ra}$. At least two of these are needed (a = 1,2,...) to explain the mass squared differences found experimentally.

%

%

%

The neutrino mass matrix is then given by

\equb \mathcal{M}_{\nu} =  -v^2\lambda_{\phi}^TM_R^{-1}\lambda_{\phi} =  -M_D^T M_R^{-1}M_D. \label{canonical} \eque
This is the see-saw result for the light neutrino masses, which are quadratic in the Dirac mass and inversely proportional to the large Majorana mass.

\subsection{The family group}

The existence of a family symmetry group $F$ (also called a horizontal symmetry) was proposed long ago \cite{Wilczek}. Under $F$, the three quark and lepton generations transform into each other. Non-Abelian finite groups have been extensively used to parametrize the mixing and mass matrices for neutrinos. Among these non-Abelian groups the tetrahedral group $A_4$ has become very popular \cite{Softly,Examples,Models,Babu,Tetrahedral,Altarelli}. The Frobenius groups which are subgroups of SU(3) also provide interesting family symmetries. As an example, the Frobenius group, $T_{7}=Z_7 \rtimes Z_3$ has been discussed in \cite{Luhn, Cao} in connection with neutrino mixing.  It has been shown in \cite{Parattu} that $T_{13}$ is suitable to produce tribimaximal mixing. In this paper, this Frobenius group will be used. It has already been described in detail in previous papers \cite{Kajiyama, Ding, Frobenius}, so only the essential characteristics of the group will be reviewed here.

$T_{13}$ is a subgroup of SU(3) generated by two elements $a$ and $b$, such that $a^{13} = I$, $b^3 = I$, and $ba = a^3b$. All elements $g$ of $T_{13}$ can be written as $g = a^mb^n$ with $0 \leq m \leq 12$ and $0 \leq n \leq 2$. Thus, the group has $13 \times 3 = 39$ elements and seven conjugacy classes given by

\equb \nonumber C_1: \{e\} \eque
\equb \nonumber C_{13}^{(1)}: \{b, ba, ba^2, \,\, ...  \,\, , ba^{11}, ba^{12}\} \eque
\equb \nonumber C_{13}^{(2)}: \{b^2, b^2a, b^2a^2, \,\, ... \,\, b^2a^{11}, b^2a^{12}\} \eque
\equb \nonumber C_{3_1}: \{a, a^3, a^9 \} \eque
\equb \nonumber C_{\bar{3}_{1}}: \{a^4, a^{10}, a^{12} \} \eque
 \equb \nonumber C_{3_2}: \{a^2, a^5, a^6 \} \eque
\equb C_{\bar{3}_{2}}: \{a^7, a^{8}, a^{11} \}. \label{classes} \eque
The group possesses three one-dimensional irreducible representations, $\bold{1}, \bold{1'}$, and $\bold{\bar{1}'}$ and four three-dimensional irreducible representations, $\bold{3_1}, \bold{\bar{3}_1}, \bold{3_2}$, and $\bold{\bar{3}_2}$. 
%
%

%
%
%


%
%

%
The Kronecker products and the Clebsch-Gordan decompositions of the representations of the group are listed in Tables \ref{table:Kron2} and \ref{table:Clebsch}, respectively. They were derived in \cite{Frobenius}, where the notation was also described.

\begin{table}[ht]
\centering      
\begin{tabular}{l}  
\hline
\hline
$\bold{1'} \otimes \bold{1'} = \bold{\bar{1}'} $ \\   
$\bold{\bar{1}'} \otimes \bold{\bar{1}'} = \bold{1'}$ \\
$\bold{1'} \otimes \bold{\bar{1}'} = \bold{1}$ \\
$\bold{3_1} \otimes \bold{3_1} = \bold{\bar{3}_1} \oplus \bold{\bar{3}_1} \oplus \bold{3_2}$   \\ 
$\bold{3_2} \otimes \bold{3_2} = \bold{\bar{3}_2} \oplus \bold{\bar{3}_1} \oplus \bold{\bar{3}_2}$   \\ 
$ \bold{3_1} \otimes \bold{\bar{3}_1} = \bold{1} \oplus \bold{1'} \oplus \bold{\bar{1}'} \oplus \bold{3_2} \oplus \bold{\bar{3}_2}$  \\
$ \bold{3_2} \otimes \bold{\bar{3}_2} = \bold{1} \oplus \bold{1'} \oplus \bold{\bar{1}'} \oplus \bold{3_1} \oplus \bold{\bar{3}_1}$  \\
$\bold{3_1} \otimes \bold{3_2} = \bold{\bar{3}_2} \oplus \bold{3_1} \oplus \bold{3_2}$   \\ 
$\bold{3_1} \otimes \bold{\bar{3}_2} = \bold{\bar{3}_2} \oplus \bold{3_1} \oplus \bold{\bar{3}_1}$   \\ 
$\bold{3_2} \otimes \bold{\bar{3}_1} = \bold{3_2} \oplus \bold{3_1} \oplus \bold{\bar{3}_1}$   \\ 
\hline     
\hline
\end{tabular} 
\caption{$T_{13} = Z_{13} \rtimes Z_3$ Kronecker products} 
\label{table:Kron2}  
\end{table} 

\begin{table}[ht]
\centering      
\begin{tabular}{l}  
\hline
\hline
\\
$ \bold{3_1} \otimes \bold{3_1} \rightarrow \left( \begin{array}{ccc}
\psi^{11} \\
\psi^{22} \\
\psi^{33} \end{array} \right)_{\bold{3_2}}, \,\,\,\,\, \left( \begin{array}{ccc}
\psi^{23} \\
\psi^{31} \\
\psi^{12} \end{array} \right)_{\bold{\bar{3}_1}}, \,\,\,\,\, \left( \begin{array}{ccc}
\psi^{32} \\
\psi^{13} \\
\psi^{21} \end{array} \right)_{\bold{\bar{3}_1}}$ \\   
\\
$\bold{3_2} \otimes \bold{3_2} \rightarrow \left( \begin{array}{ccc}
\psi^{22} \\
\psi^{33} \\
\psi^{11} \end{array} \right)_{\bold{\bar{3}_1}}, \,\,\,\,\, \left( \begin{array}{ccc}
\psi^{23} \\
\psi^{31} \\
\psi^{12} \end{array} \right)_{\bold{\bar{3}_2}}, \,\,\,\,\, \left( \begin{array}{ccc}
\psi^{32} \\
\psi^{13} \\
\psi^{21} \end{array} \right)_{\bold{\bar{3}_2}}$ \\
\\
$ \bold{3_1} \otimes \bold{\bar{3}_1} \rightarrow   \left( \begin{array}{ccc}
\psi^1_2 \\
\psi^2_3 \\
\psi^3_1 \end{array} \right)_{\bold{\bar{3}_2}},  \,\,\,\,\,  \left( \begin{array}{ccc}
\psi^2_1 \\
\psi^3_2 \\
\psi^1_3 \end{array} \right)_{\bold{3_2}}, \,\,\,\,\, (\psi^1_1+ \psi^2_2 + \psi^3_3)_{\bold{1}},$ \\
\\
$ \,\,\,\,\, \,\,\,\,\, \,\,\,\,\, \,\,\,\,\, \,\,\,\,\, \,\,\,\,\, (\psi^1_1+ \omega \psi^2_2 + \omega^2 \psi^3_3)_{\bold{1'}} , \,\,\,\,\,
 (\psi^1_1 + \omega^2 \psi^2_2 + \omega \psi^3_3)_{\bold{\bar{1}'}}$ \\
 \\
$ \bold{3_2} \otimes \bold{\bar{3}_2} \rightarrow   \left( \begin{array}{ccc}
\psi^3_2 \\
\psi^1_3 \\
\psi^2_1 \end{array} \right)_{\bold{\bar{3}_1}},  \,\,\,\,\,  \left( \begin{array}{ccc}
\psi^2_3 \\
\psi^3_1 \\
\psi^1_2 \end{array} \right)_{\bold{3_1}}, \,\,\,\,\, (\psi^1_1+ \psi^2_2 + \psi^3_3)_{\bold{1}},$ \\
\\ 
$  \,\,\,\,\,  \,\,\,\,\,  \,\,\,\,\,  \,\,\,\,\,  \,\,\,\,\,  \,\,\,\,\, (\psi^1_1+ \omega \psi^2_2 + \omega^2 \psi^3_3)_{\bold{1'}},  \,\,\,\,\, 
 (\psi^1_1 + \omega^2 \psi^2_2 + \omega \psi^3_3)_{\bold{\bar{1}'}}$ \\
 \\
$ \bold{3_1} \otimes \bold{3_2} \rightarrow \left( \begin{array}{ccc}
\psi^{33} \\
\psi^{11} \\
\psi^{22} \end{array} \right)_{\bold{3_1}}, \,\,\,\,\, \left( \begin{array}{ccc}
\psi^{31} \\
\psi^{12} \\
\psi^{23} \end{array} \right)_{\bold{\bar{3}_2}}, \,\,\,\,\,\left( \begin{array}{ccc}
\psi^{32} \\
\psi^{13} \\
\psi^{21} \end{array} \right)_{\bold{3_2}} $ \\
\\
$ \bold{3_1} \otimes \bold{\bar{3}_2} \rightarrow \left( \begin{array}{ccc}
\psi^1_1 \\
\psi^2_2 \\
\psi^3_3 \end{array} \right)_{\bold{\bar{3}_1}}, \,\,\,\,\,
 \left( \begin{array}{ccc}
\psi^2_3 \\
\psi^3_1 \\
\psi^1_2 \end{array} \right)_{\bold{\bar{3}_2}}, \,\,\,\,\,
 \left( \begin{array}{ccc}
\psi^2_1 \\
\psi^3_2 \\
\psi^1_3 \end{array} \right)_{\bold{3_1}}$   \\ 
\\
$\bold{3_2} \otimes \bold{\bar{3}_1} \rightarrow  \left( \begin{array}{ccc}
\psi^1_1 \\
\psi^2_2 \\
\psi^3_3 \end{array} \right)_{\bold{3_1}}, \,\,\,\,\,
 \left( \begin{array}{ccc}
\psi^1_2 \\
\psi^2_3 \\
\psi^3_1 \end{array} \right)_{\bold{\bar{3}_1}}, \,\,\,\,\,
 \left( \begin{array}{ccc}
\psi^3_2 \\
\psi^1_3 \\
\psi^2_1 \end{array} \right)_{\bold{3_2}} $   \\ 
\\
\hline     
\hline
\end{tabular} 
\caption{$Z_{13} \rtimes Z_3$ Clebsch-Gordan decompositions.} 
\label{table:Clebsch}  
\end{table} 
\subsection{The mixing matrix}

The charged lepton masses are generated by the dimension-4 operator:

\equb \mathcal{O}_4 = \bar{\psi}_L\phi l_R, \eque
with $\phi$ the Higgs doublet, $\psi$ the lepton doublet, and $l_R$ the charge conjugated lepton. The mass terms for the neutrinos $\nu$ and charged leptons $l$ in the Lagrangian can be written as

\equb \mathcal{L}_{mass} = -\bar{l}_{L\alpha}(M_{l})_{\alpha \beta}l_{R\beta} - \frac{1}{2}\nu_{L\alpha}^T(M_{\nu})_{\alpha\beta}C\nu_{L \beta} + h.c, \eque
where $\alpha, \beta$ denote the flavor indices $e, \mu, \tau$. The masses are then obtained by finding the eigenvalues of the mass matrices through diagonalization: 

\equb U_{L}^{\dagger}M_{l}U_R \equiv D_l =   \left(\begin{array}{ccc} m_{e} & 0 & 0  \\
0& m_{\mu}& 0\\
0 &0 & m_{\tau} \end{array} \right), \label{mixing} \eque
\equb U_{\nu}^TM_{\nu}U_{\nu} \equiv D_{\nu} =  \left(\begin{array}{ccc} m_1 & 0 & 0  \\
0& m_2& 0\\
0 &0 & m_3 \end{array} \right).\eque
The left-handed charged lepton mass basis is given by

\equb l_{L\alpha} \equiv (U_L)_{\alpha i} l_{Li}, \eque
and the neutrino mass basis by

\equb \nu_{\alpha} \equiv (U_{\nu})_{\alpha i} \nu_{i}, \eque
with $i = 1,2,3$ as in (\ref{mix}), so that

\equb \left(\begin{array}{c}\nu_{\alpha}\\
l_{L\alpha} \end{array} \right) = \left(\begin{array}{c}U_{\nu} \nu_{i}\\
U_L l_{Li} \end{array} \right) = U_L \left(\begin{array}{c}U_L^{-1}U_{\nu} \nu_{i}\\
l_{Li} \end{array} \right). \eque
Thus, the lepton mixing matrix is given by $U = U_{L}^{\dagger}U_{\nu}$.

The neutrino mass matrix results from integrating out the heavy right-handed neutrinos as mentioned. The Lagrangian containing the Majorana mass term for the right-handed neutrinos and the Dirac mass term coupling the left- and right-handed neutrinos can be diagonalized in the same way: 

\equb U_{DL}^{\dagger}M_{D}U_{DR} \equiv D_D =   \left(\begin{array}{ccc} m_{1} & 0 & 0  \\
0& m_{2}& 0\\
0 &0 & m_{3} \end{array} \right)_D, \label{UDL} \eque
\equb U_{R}^TM_{R}U_{R} \equiv D_{R} =  \left(\begin{array}{ccc} m_1 & 0 & 0  \\
0& m_2& 0\\
0 &0 & m_3 \end{array} \right)_R. \label{UR}\eque
This means that the neutrino mass matrix can be written in terms of these unitary matrices and the diagonal Dirac and right-handed mass matrices,
     
\equb M_{\nu} = - M_D^TM_R^{-1}M_D = - U_{DR}^*D_DU_{DL}^TU_RD_RU_R^TU_{DL}D_DU_{DR}^{\dagger}, \eque
and be diagonalized as seen above. 

\section{Model building}

Given the review of the canonical see-saw mechanism in section 2.2 and the group theory in section 2.3, a model can now be built by implementing the family symmetry group $T_{13}=Z_{13} \rtimes Z_3$ and the type I see-saw mechanism with neutrino mixing. As in the standard procedure, the lepton doublet fields $\psi_a$ ($a = 1,2,3$), the right-handed lepton singlet fields $l_{Ra}$, the right-handed neutrino singlet fields $N_{Ra}$ and the scalar fields are assigned to various representations of $T_{13}$. The invariants in the Lagrangian are found. After the scalar fields are allowed to acquire their vacuum expectation values, the resulting right-handed neutrino mass matrix $M_R$, the charged lepton mass matrix $M_l$, the Dirac mass matrix $M_D$, and the neutrino mass matrix $M_{\nu}$ are diagonalized, whereby the mixing matrix is obtained as outlined earlier.

\subsection{The see-saw mechanism and the Frobenius group $T_{13}$}

In the following, the application of the group $T_{13}$ to the canonical see-saw mechanism will be studied, with three right-handed neutrinos and three left-handed neutrinos. The Lagrangian terms of importance are the following:

\equb \mathcal{L}^{\nu} = \frac{1}{2}N_R^TC^{-1}MN_R + \frac{1}{2}\lambda_{\chi}N_R^TC^{-1}N_R\chi + \frac{1}{2}\lambda_{\xi}N_R^TC^{-1}N_R\xi + \lambda_{\phi}\bar{\psi}_L\phi N_R + \bar{\psi}_L \phi l_R, \eque
where an extra singlet - $\xi$ - has been introduced \cite{Porto}.\footnote{A single scalar singlet, $\chi$, is not sufficient to give solutions to this coupling between the Frobenius group and the canonical see-saw mechanism.} The right-handed neutrinos are assigned to three-dimensional representations.\footnote{An analysis has been carried out showing no solutions when placing the $N_R$'s in the one-dimensional representations, when fitting with tribimaximal mixing.} Looking at the Clebsch-Gordan decompositions, group theory excludes the plain mass term $N_R^TC^{-1}MN_R$ for this choice. In Table \ref{RH} the possible representations of the right-handed neutrinos and the scalar singlets $\chi$ and $\xi$ in the terms $\lambda_{\chi}N_R^TC^{-1}N_R\chi$ and $\lambda_{\xi}N_R^TC^{-1}N_R\xi$ are listed. 

\begin{table}[h!]
\centering      
\begin{tabular}{c | c | c | c}  
\hline\hline                        
Case &$N_R$ & $\chi$ &  $\xi$ \\  
\hline                    
1 & $\bold{3_1}$ & $\bold{3_1}$ & $\bold{\bar{3}_2}$  \\   
2 & $\bold{3_2}$ &$\bold{3_2}$ & $\bold{3_1}$ \\

\hline     
\end{tabular} 
\caption{Possible representations for the three fields in the right-handed neutrino mass term.} 
\label{RH}  
\end{table} 
When the family symmetry is broken, the scalar singlets acquire the vacuum expectation values $\langle \chi \rangle$ and $\langle \xi \rangle$. The right-handed neutrino mass matrices are then given by

 \equb \text{Case} \, 1: \,\,\,\,\, M_R = \left( \begin{array}{ccc}
\lambda_{\xi}\langle \xi_1 \rangle & \lambda_{\chi}\langle \chi_3 \rangle & \lambda_{\chi}\langle \chi_2 \rangle \\
\lambda_{\chi}\langle  \chi_3 \rangle & \lambda_{\xi}\langle \xi_2 \rangle & \lambda_{\chi}\langle \chi_1 \rangle \\
\lambda_{\chi}\langle \chi_2 \rangle & \lambda_{\chi}\langle \chi_1 \rangle & \lambda_{\xi}\langle \xi_3 \rangle \end{array} \right),  \label{1} \eque
 \equb \text{Case} \, 2: \,\,\,\,\, M_R = \left( \begin{array}{ccc}
\lambda_{\xi}\langle \xi_3 \rangle & \lambda_{\chi}\langle \chi_3 \rangle & \lambda_{\chi}\langle \chi_2 \rangle \\
\lambda_{\chi}\langle  \chi_3 \rangle & \lambda_{\xi}\langle \xi_1 \rangle & \lambda_{\chi}\langle \chi_1 \rangle \\
\lambda_{\chi}\langle \chi_2 \rangle & \lambda_{\chi}\langle \chi_1 \rangle & \lambda_{\xi}\langle \xi_2 \rangle \end{array} \right), \label{2}\eque
The inverse right-handed neutrino mass matrices are then given by:
\\
\\
Case 1:
\equb M_R^{-1} \sim \left( \begin{array}{ccc}
-\lambda_{\chi}^2\langle\chi_1\rangle^2 + \lambda_{\xi}^2\langle\xi_2\rangle\langle\xi_3\rangle & \lambda_{\chi}^2\langle\chi_1\rangle\langle\chi_2\rangle-\lambda_{\xi}\lambda_{\chi}\langle\chi_3\rangle\langle\xi_3\rangle & \lambda_{\chi}^2\langle\chi_1\rangle\langle\chi_3\rangle-\lambda_{\xi}\lambda_{\chi}\langle\chi_2\rangle\langle\xi_2\rangle \\
\lambda_{\chi}^2\langle\chi_1\rangle\langle\chi_2\rangle-\lambda_{\xi}\lambda_{\chi}\langle\chi_3\rangle\langle\xi_3\rangle & -\lambda_{\chi}^2\langle\chi_2\rangle^2 + \lambda_{\xi}^2\langle\xi_1\rangle\langle\xi_3\rangle & \lambda_{\chi}^2\langle\chi_2\rangle\langle\chi_3\rangle-\lambda_{\xi}\lambda_{\chi}\langle\chi_1\rangle\langle\xi_1\rangle \\
\lambda_{\chi}^2\langle\chi_1\rangle\langle\chi_3\rangle-\lambda_{\xi}\lambda_{\chi}\langle\chi_2\rangle\langle\xi_2\rangle & \lambda_{\chi}^2\langle\chi_2\rangle\langle\chi_3\rangle-\lambda_{\xi}\lambda_{\chi}\langle\chi_1\rangle\langle\xi_1\rangle & -\lambda_{\chi}^2\langle\chi_3\rangle^2 + \lambda_{\xi}^2\langle\xi_1\rangle\langle\xi_2\rangle \end{array} \right),\eque
Case 2:
\equb  M_R^{-1} \sim \left( \begin{array}{ccc}
-\lambda_{\chi}^2\langle\chi_1\rangle^2 + \lambda_{\xi}^2\langle\xi_1\rangle\langle\xi_2\rangle & \lambda_{\chi}^2\langle\chi_1\rangle\langle\chi_2\rangle-\lambda_{\xi}\lambda_{\chi}\langle\chi_3\rangle\langle\xi_2\rangle & \lambda_{\chi}^2\langle\chi_1\rangle\langle\chi_3\rangle-\lambda_{\xi}\lambda_{\chi}\langle\chi_2\rangle\langle\xi_1\rangle \\
\lambda_{\chi}^2\langle\chi_1\rangle\langle\chi_2\rangle-\lambda_{\xi}\lambda_{\chi}\langle\chi_3\rangle\langle\xi_2\rangle & -\lambda_{\chi}^2\langle\chi_2\rangle^2 + \lambda_{\xi}^2\langle\xi_3\rangle\langle\xi_2\rangle & \lambda_{\chi}^2\langle\chi_2\rangle\langle\chi_3\rangle-\lambda_{\xi}\lambda_{\chi}\langle\chi_1\rangle\langle\xi_3\rangle \\
\lambda_{\chi}^2\langle\chi_1\rangle\langle\chi_3\rangle-\lambda_{\xi}\lambda_{\chi}\langle\chi_2\rangle\langle\xi_1\rangle & \lambda_{\chi}^2\langle\chi_2\rangle\langle\chi_3\rangle-\lambda_{\xi}\lambda_{\chi}\langle\chi_1\rangle\langle\xi_3\rangle & -\lambda_{\chi}^2\langle\chi_3\rangle^2 + \lambda_{\xi}^2\langle\xi_3\rangle\langle\xi_1\rangle \end{array} \right), \eque
up to an irrelevant normalization factor. As seen from (\ref{1}) and (\ref{2}) the only difference lies in the vacuum alignment of the $\xi$. 

In the dimension 4 operator, the assignments of the fields to the irreducible representations of $T_{13}$ have already been investigated in \cite{Frobenius}. Therefore, the solutions will just be listed here. When placing the fields in various representations, it was found that only six cases led to useful solutions. These representations and the charged lepton mixing matrices, $U_L^{\dagger}$, corresponding to these six cases can be seen in Table \ref{charged}. The charge conjugations of these representations lead to the same results. 

\begin{table}[h!]
\centering      
\begin{tabular}{c |c c c | c}  
\hline\hline                        
Case &$\bar{\psi}_L$ & $\phi$ & $l_R$ & $U_L^{\dagger}$ \\  
\hline                    
& & & & \\
\multirow{4}{*}{A}& $\bold{3_1}$ & $\bold{3_1}$ & $\bold{3_1}$   & \multirow{4}{*} {$\frac{1}{\sqrt{3}}\left( \begin{array}{ccc}
 1 & 1& 1 \\
 \omega &1& \omega^2\\
 \omega^2 & 1 & \omega \end{array} \right)$}  \\   
& $\bold{3_2}$ &$\bold{3_2}$ &$\bold{3_2}$ & \\
& $\bold{3_1}$ & $\bold{\bar{3}_1}$& $\bold{1}, \bold{1'}, \bold{\bar{1}'}$ & \\
& $\bold{3_2}$ &$\bold{\bar{3}_2}$ & $\bold{1}, \bold{1'}, \bold{\bar{1}'}$  &\\
& & & & \\
\hline
& & & & \\
\multirow{2}{*}{B} & $\bold{3_1}$ & $\bold{1}, \bold{1'}, \bold{\bar{1}'} $& $\bold{\bar{3}_1}$ &\multirow{2}{*} {$\left( \begin{array}{ccc}
1 & 0& 0 \\
0 &1& 0\\
0 & 0 & 1 \end{array} \right)$} \\
& $\bold{3_2}$& $\bold{1}, \bold{1'}, \bold{\bar{1}'}$ & $\bold{\bar{3}_2}$ &\\
& & & & \\
& & & & \\

\hline     
\end{tabular} 
\caption{Possible representations for the three fields in the dimension 4 operator, and the corresponding charged lepton mixing matrix.} 
\label{charged}  
\end{table} 

When using tribimaximal mixing as a guideline, the neutrino mixing matrix $U_{\nu}$ can be found from the charged lepton mixing matrix $U_L^{\dagger}$ through the following relation:

\equb U_{TB} = U_L^{\dagger}U_{\nu}. \eque
The neutrino mixing matrices must, therefore, in cases A and B of Table \ref{charged} be given by

\equb U_{\nu A} = \frac{1}{\sqrt{2}} \left( \begin{array}{ccc}
1 & 0& -1 \\
0 &\sqrt{2}& 0\\
1 & 0 & 1 \end{array} \right), \eque
and 
\equb U_{\nu B} = U_{TB} = \left( \begin{array}{ccc}
\frac{2}{\sqrt{6}} & \frac{1}{\sqrt{3}} & 0 \\
-\frac{1}{\sqrt{6}} & \frac{1}{\sqrt{3}} & \frac{1}{\sqrt{2}} \\
-\frac{1}{\sqrt{6}} & \frac{1}{\sqrt{3}} & -\frac{1}{\sqrt{2}} \end{array} \right), \eque
respectively, corresponding to the two possible charged lepton mixing matrices, in order to get tribimaximal mixing. The corresponding mass matrices have the forms

\equb M_{\nu A} \sim \left( \begin{array}{ccc}
\alpha & 0 & \beta \\
0 &\gamma& 0 \\
\beta & 0 & \alpha \end{array} \right), \label{mass} \eque
and

\equb M_{\nu B}  \sim  \left( \begin{array}{ccc}
\gamma & \beta & \beta  \\
\beta & \alpha & \delta \\
\beta & \delta & \alpha \end{array} 
\right). \label{trib} \eque
The last matrix needs to satisfy the condition $\delta = \gamma + \beta - \alpha$ to result in tribimaximal mixing.

Taking into account the possible representations of $N_R$, $\bar{\psi}_L$, and $\phi$ as in Tables \ref{RH} and \ref{charged}, a detailed analytical analysis shows that the Dirac mass matrix can acquire three forms that lead to tribimaximal mixing, namely,\footnote{As an example, one can have the fields in the representations as seen in the last line of case A in table \ref{charged}, where $\bar{\Psi}_L \sim \bold{3_2}$ and $\phi \sim \bold{\bar{3}_2}$. This leads to the combinations $\{\bar{\psi}_{L3}\phi_2, \bar{\psi}_{L1}\phi_3, \bar{\psi}_{L2}\phi_1\}_{\bold{\bar{3}_1}}$ and $\{\bar{\psi}_{L2}\phi_3,\bar{\psi}_{L3}\phi_1,\bar{\psi}_{L1}\phi_2\}_{\bold{3_1}}$ as follows from table \ref{table:Clebsch}. These can then be connected with the $N_R$ in a $\bold{3_1}$ and a $\bold{\bar{3}_1}$ representation, respectively, leading to the two off-diagonal forms. Continuing systematically like this, one finds that only four forms are possible. Of these, only the three listed above will lead to tribimaximal mixing.}

\equb M_D \sim I, \,\,\,\,\, M_D \sim  \left( \begin{array}{ccc}
0 & X & 0 \\
0 & 0 & X \\
X & 0 & 0 \end{array} \right), \,\,\,\,\, M_D \sim  \left( \begin{array}{ccc}
0 & 0 & X \\
X & 0 & 0 \\
0 & X & 0 \end{array} \right). \eque
The Higgs doublet $\phi$ is aligned in the (1,1,1) direction, when placed in a three dimensional representation, to acquire nondegenerate masses for the charged leptons in the dimension 4 operator. The Dirac mass matrices will lead to the same neutrino mass matrices, since the last two will only perturb the vacuum alignments in the case where the Dirac mass matrix is diagonal. The shapes can be obtained by choosing representations for $\bar{\psi}_L$ and $\phi$ in the term $\lambda_{\phi}\bar{\psi}_LN_R\phi$ in various ways. The six possibilities from the dimension 4 operator need to be taken into consideration when making these choices. It turns out that only cases where the Higgs doublet is placed in a three dimensional representation are possible. Therefore, case B of Table \ref{charged} can be excluded. All the possible ways of placing the various fields in different representations are assembled in Table \ref{summ}.

\begin{table}[h!]
\centering      
\begin{tabular}{c c c c c c}  
\hline\hline                        
$M_D$ & $N_R$ & $\chi$ & $\xi$ & $\bar{\psi}_L$ & $\phi$ \\  
\hline                    
 \multirow{2}{*}{Diagonal}  & $\bold{3_2}$ &$\bold{3_2}$ & $\bold{3_1}$  & $\bold{3_1}$ & $\bold{\bar{3}_1}$\\
 & $\bold{3_2}$ &$\bold{3_2}$ & $\bold{3_1}$  & $\bold{\bar{3}_1}$ & $\bold{\bar{3}_1}$\\
 \hline
 \multirow{4}{*}{Off-diagonal} & $\bold{3_1}$  & $\bold{3_1}$ & $\bold{\bar{3}_2}$   & $\bold{3_2}$  & $\bold{\bar{3}_2}$\\  
 & $\bold{3_1}$  & $\bold{3_1}$ & $\bold{\bar{3}_2}$   & $\bold{3_2}$  & $\bold{3_2}$\\   
& $\bold{3_1}$  & $\bold{3_1}$ & $\bold{\bar{3}_2}$   & $\bold{\bar{3}_2}$  & $\bold{3_2}$\\  
 & $\bold{3_2}$ &$\bold{3_2}$ & $\bold{3_1}$  & $\bold{\bar{3}_1}$ & $\bold{3_1}$\\
\hline     
\end{tabular} 
\caption{Possible representations for the five fields and the corresponding form of the Dirac mass matrix.} 
\label{summ}  
\end{table} 

In the two possibilities where $M_D$ is diagonal, the neutrino mass matrix will have the same structure as $M_R^{-1}$. In the four possibilities where $M_D$ is off-diagonal, the neutrino mass matrix can be found from (\ref{canonical}) to be

\equb M_{\nu} \sim  \langle\phi\rangle ^2 \left( \begin{array}{ccc}
0 & 1& 0 \\
0 & 0 & 1 \\
 1& 0 & 0 \end{array} \right) M_R^{-1} \left( \begin{array}{ccc}
0 & 0 & 1 \\
1 & 0 & 0 \\
0 & 1 & 0 \end{array} \right). \eque
The form of $M_{\nu A}$ can be obtained by choosing the following vacuum expectation values for the neutral component of the two scalar singlets and the Higgs doublet in the diagonal case,

\equb \langle \phi \rangle = v  \left( \begin{array}{c}
1 \\
1 \\
1 \end{array} \right), \,\,\,\,\, \langle \chi \rangle = u_1 \left( \begin{array}{c}
0 \\
1 \\
0 \end{array} \right), \,\,\,\,\, \langle \xi \rangle = u_2 \left( \begin{array}{c}
1 \\
1 \\
1 \end{array} \right), \eque
and in the off-diagonal case,

\equb \langle \phi \rangle = v  \left( \begin{array}{c}
1 \\
1 \\
1 \end{array} \right), \,\,\,\,\, \langle \chi \rangle = u_1 \left( \begin{array}{c}
0 \\
0 \\
1 \end{array} \right), \,\,\,\,\, \langle \xi \rangle = u_2 \left( \begin{array}{c}
1 \\
1 \\
1 \end{array} \right). \eque
After symmetry breaking with these vacuum alignments, the neutrino mass matrix will, in both cases, be given by

\equb M_{\nu} = - M_D^TM_R^{-1}M_D = \frac{1}{\lambda_{\xi}^3u_2^3-\lambda_{\xi}\lambda_{\chi}^2u_2u_1^2} \left( \begin{array}{ccc}
\lambda_{\xi}^2u_2^2&0  &  -\lambda_{\xi}\lambda_{\chi}u_1u_2 \\
0 &  -\lambda_{\chi}^2u_1^2 + \lambda_{\xi}^2u_2^2&  0 \\
-\lambda_{\xi}\lambda_{\chi}u_1u_2 & 0 & \lambda_{\xi}^2u_2^2 \end{array} \right), \eque
which has the desired shape of (\ref{mass}). Note, however, the constraint $\gamma = \alpha - \frac{\beta^2}{\alpha}$, so that the form of the mass matrix is now more constrained:

\equb M_{\nu A} \sim \left( \begin{array}{ccc}
\alpha & 0 & \beta \\
0 &\alpha-\frac{\beta^2}{\alpha}& 0 \\
\beta & 0 & \alpha \end{array} \right). \eque
The possible choices for assigning representations to the various fields have now been exhausted leading to only a single possible neutrino mass matrix. The result can now be explored.

\section{Predictions}

The result of this paper is that the neutrino mass matrix can have a single shape that is highly constrained, and depends on only two parameters:

\equb M_{\nu A} \sim \left( \begin{array}{ccc}
\alpha & 0 & \beta \\
0 &\alpha-\frac{\beta^2}{\alpha}& 0 \\
\beta & 0 & \alpha \end{array} \right). \eque
Taking into account the experimental knowledge about neutrino masses, the model allows some predictions. The eigenvalues of the mass matrix are given by

\equb m_1 = \alpha + \beta, \,\,\,\,\, m_2 = \alpha- \frac{\beta^2}{\alpha}, \,\,\,\,\, m_3 = \alpha - \beta. \eque
The mass squared differences can be found in terms of $\alpha$ and $\beta$:

\equb  \Delta m_{12}^2 = |m_2|^2-|m_1|^2 = |\alpha-\frac{\beta^2}{\alpha}|^2-|\alpha+\beta|^2, \label{S}\eque
\equb  \Delta m_{31}^2 = |m_3|^2-|m_1|^2  = |\alpha-\beta|^2-|\alpha + \beta|^2. \label{A}\eque
At best fit, the experimental values are

\equb\Delta m_{12}^2 = 7.59^{+0.20}_{-0.18}  \cdot 10^{-5} eV^2,  \eque
\equb \Delta m_{31}^2 =  2.50^{+0.09}_{-0.16}  \cdot 10^{-3} eV^2 \, (\text{normal hierarchy}). \eque
\equb \Delta m_{31}^2 =  -(2.40^{+0.08}_{-0.09})  \cdot 10^{-3} eV^2 \, (\text{inverted hierarchy}). \eque
The two equations (\ref{S}) and (\ref{A}) can be solved for $\alpha$ and $\beta$. In the normal hierarchy, $|m_3| \gg |m_1| \simeq |m_2|$. In the inverted hierarchy, $|m_3| \ll |m_1| \sim |m_2|$. 

Since this paper assumes no CP violation, using tribimaximal mixing as a first order approximation, $\alpha$ and $\beta$ must also be real to first order. The solution for these as well as the neutrino mass eigenvalues can be seen in Table \ref{predictions} \footnote{In order to acquire positive masses for these Majorana neutrinos, it is possible to redefine the Majorana fields $\chi_n = \Sigma_a (U_{na}\nu_{La} + \eta_n U_{na}\nu_{Ra}^c)$, where $\eta_n= \pm 1$ are CP eigenvalues, $U_{na}$ is the orthogonal matrix that diagonalizes the real neutrino mass matrix and we have $(CP)\nu_{La}(CP)^{-1} = \nu_{Ra}^c$. Therefore, $UM_{\nu}U^T = \eta_nm_n$, and the masses $m_n$ are positive \cite{Wolfenstein2}. The solutions that differ only by a sign can therefore be omitted.}. 

\begin{table}[h!]
\centering      
\begin{tabular}{c | c c | c c  }  
\hline\hline                        
Hierarchy & \multicolumn{2}{c|}{Normal}  &  \multicolumn{1}{c}{Inverted}    \\
\hline
 $\alpha \, [10^{-2} eV]$ & $2.85^{+0.04}_{-0.09}$   & $2.32^{+0.04}_{-0.08}$  & $1.70^{+0.03}_{-0.04}$  \\   
 $\beta \, [10^{-2} eV]$ & $-(2.26^{+0.05}_{-0.09}) $& $-(2.77^{+0.05}_{-0.09})$ & $3.42^{+0.06}_{-0.07}$  \\ 
 $m_1 \, [10^{-2} eV]$ & $-(0.59^{+0.13}_{-0.14})$ & $0.45^{+0.13}_{-0.13}$ & $5.12^{+0.09}_{-0.11}$ \\ 
 $m_2 \, [ 10^{-2} eV]$ &  $-(1.06^{+0.20}_{-0.23})$ & $0.99^{+0.32}_{-0.31}$ & $-(5.18^{+0.46}_{-0.42})$  \\
 $m_3 \, [10^{-2} eV]$ & $-(5.11^{+0.09}_{-0.18})$& $-(5.09^{+0.09}_{-0.17})$ & $-(1.72^{+0.10}_{-0.10})$ \\ 
\hline     
\end{tabular} 
\caption{Predictions for neutrino mass eigenvalues using best fit values of the mass squared differences.} 
\label{predictions}  
\end{table} 


From these values, the possible regime for neutrinoless double beta decay can be explored. The rate of this decay is proportional to the modulus of the (ee) entry of the effective neutrino mass matrix, given by

\equb |m_{\text{ee}}| = |m_1U_{11}^2+m_2U_{12}^2+m_3U_{13}^2|. \eque
The (ee) entry of the neutrino mass matrix with tribimaximal mixing and CP conservation is then given by:

\equb |m_{\text{ee}}| = \frac{1}{3}|2m_1+m_2|. \eque
The absolute value of the (ee) entry corresponding to the three possible values of the neutrino mass eigenvalues can be seen in Table \ref{mee}. The upper limit for this entry today is given by $0.38 eV$. These predictions are well below this limit. Furthermore, this model can be tested in the near future, since upcoming neutrinoless double beta decay experiments will have sensitivities down to the order of $10^{-2} eV$ \cite{Aalseth,Abt}. Therefore, this model could soon be confronted with experiment. 

\begin{table}[h!]
\centering      
\begin{tabular}{c | c c | c c c | c}  
\hline\hline                        
Hierarchy  & $|m_{ee}| [10^{-2} eV]$\\  
\hline                    
\multirow{2}{*}{Normal}   & $0.75^{+0.15}_{-0.17}$ \\   
  & $0.63^{+0.19}_{-0.19}$\\  
\hline
\multirow{1}{*}{Inverted}  & $1.69^{+0.20}_{-0.23}$ \\
\hline     
\end{tabular} 
\caption{Predictions for the modulus of the (ee) entry of the neutrino mass matrix.} 
\label{mee}  
\end{table} 

If these results survive future experiments, e.g. neutrinoless double beta decay, the model of combining the Frobenius group $T_{13}$ with the canonical see-saw mechanism can be seen as a good candidate to predict neutrino masses. 

\subsection{Assumptions}

In order to constrain the systematic search, the following assumptions have been used during the model building:
\\
\\
i) The vacuum expectation value of the Higgs doublet in the dimension 4 operator and the Dirac mass term was aligned in the (1,1,1) direction.
\\
\\
ii) No coupling was assumed between the scalar fields in the charged lepton sector and the neutrino sector, that is, the sequestering problem has not been addressed. 
\\
\\
iii) Since tribimaximal mixing gives a fairly good description of the experimental mixing matrix, at least as a first approximation, it was used as a phenomenological guideline. 
\\
\\
iv) It was assumed that the region of the parameter space of the scalar potential where the heavy scalar singlets develop vacuum expectation values lies in a different region than at the scale where the Higgs doublet acquires its vacuum expectation values. 

\section{Discussion and conclusion}

There is a very nice compatibility between the Frobenius group $T_{13}$ and the canonical see-saw mechanism when applied to neutrino mixing, using tribimaximal mixing as a guideline. It is also interesting to see that the Frobenius group does not allow a bare mass term for the right-handed neutrinos, but only a Yukawa coupling of these with two scalar singlets. Furthermore, the same Higgs doublets can be used in both the charged lepton mass term and the Dirac mass term. 

The assumptions allow only six possibilities for the representations of the fields. Furthermore, the constraints from applying both the family symmetry group $T_{13}$ and the canonical see-saw mechanism narrow down the parameters in the neutrino mass matrix enough to make predictions. Using experimental data, we have predicted the physical neutrino masses. Moreover, we have predicted values for the $|m_{ee}|$ parameter which will be tested in future neutrinoless double beta decay experiments. In the inverted hierarchy case, the $|m_{ee}|$ parameter is almost a magnitude higher than in the normal case. If this is the case, the next neutrinoless double beta decay experiments may be able to reveal information about the mass hierarchy as well. 

In this paper, the sequestering problem has not been addressed. When the Higgs of the dimension 4 operator acquires a vev in the (1,1,1) direction, the $T_{13}$ symmetry is broken down to $Z_3$. As the various scalar fields of the dimension 5 operators acquire vacuum expectation values, the symmetry is also broken down to $Z_2$. The $Z_2$ subgroup does not commute with the $Z_3$ subgroup, and the vacuum alignments cannot be sequestered from each other unless the interaction terms in the Higgs potential which connect these fields vanish \cite{Keum}. Tribimaximal mixing is therefore a lowest order approach, and we need to study allowed deviations within the $T_{13}$ structure. Certain articles have already described modifications to tribimaximal mixing \cite{Minimal} \cite{Harrison}. These could be used with this theory.  Also, CP violation could be considered. 

It would also be interesting to explore the compatibility of this group with the other see-saw mechanisms that have been addressed in the literature \cite{Schechter, Valle}.
\\
\\
$\bold{Acknowledgements}$
\\
\\
The author wishes to thank A. Zee and P. H. Damgaard for helpful discussions. She would also like to thank D. O'Connell for careful reading of the manuscript.


\end{document}